\documentclass[extra,mreferee]{gji}

\usepackage[pdftex]{graphicx}
\usepackage{epstopdf}
\usepackage{nicefrac}
\usepackage{amsmath}
\usepackage{amssymb}
\usepackage{placeins}
\usepackage{booktabs}
\usepackage{verbatim}
\usepackage[usenames, dvipsnames]{color}
\newcommand{\mbf}[1]{\mathbf{#1}}
\usepackage{newfloat}
\usepackage{lscape}
\usepackage{multirow}
\usepackage{soul}
\usepackage{color}
\sethlcolor{yellow}
\usepackage[T1]{fontenc}

\title[Probabilistic multiphysics inversion with variable complexity]{Multiphysics inversion with variable complexity of receiver-function, surface-wave dispersion and magnetotelluric data reduces uncertainty for lithosphere structure}

\author[Shahsavari et al.]{P. Shahsavari$^1$, J. Dettmer$^{1,2}$, M.J. Unsworth$^3$ and A. Schaeffer$^4$ \\
  $^1$Department of Earth, Energy, and Environment, University of Calgary, Calgary, Alberta, Canada \\
  $^2$Yukon Geologic Survey, Whitehorse, Yukon, Canada \\
  $^3$Department of Physics, University of Alberta, Edmonton, Alberta, Canada \\
  $^4$Geological Survey of Canada, Pacific Division, Natural Resources Canada, Sidney, BC, Canada}

\pagerange{\pageref{firstpage}--\pageref{lastpage}}

\begin{document}

\label{firstpage}

\maketitle

\begin{summary}
Probabilistic multiphysics inversion is the process of inferring parameter values for Earth models by fitting the data from multiple but distinct physical processes that may contain complementary information. For example, magnetotelluric data often resolve the lithosphere-asthenosphere boundary beneath cratons because of their sensitivity to change in resistivity associated with partial melting. However, they are not sensitive to the Moho because there is not a significant change in resistivity. In contrast, seismic data resolve the Moho boundary consistently but lack resolution of lithosphere-asthenosphere boundary beneath cratons due to a small negative velocity contrast commonly caused by partial melting. Using these complementary qualities has the potential to reduce the uncertainty in the resulting Earth models, but existing methods often rely on subjective assumptions and have not fully developed. Robust uncertainty quantification requires parameterizations that can accommodate various parameter types and can combine information from distinct physical signals. We present a probabilistic multiphysics inversion based on Bayesian inference with trans-dimensional models that has these desired properties. We jointly consider magnetotelluric, receiver function, and Rayleigh-wave dispersion data to infer one-dimensional lithospheric structure in the vicinity of Athabasca, Canada. The location is on the North American Craton with a cover of sediments from the Western Canada Sedimentary Basin. 
The trans-dimensional model uses layer nodes that include parameters that are activated or deactivated based on data information. Furthermore, the number of nodes is based on data information. Hence, the parameterization uncertainty is included in the uncertainty estimates. Furthermore, the layer nodes permit trans-dimensional decoupling such that some discontinuities may be represented by only some of the parameters. In probabilistic multiphysics inversion, it is important that the various data types are weighed objectively. Here, the weights are the data covariance matrices of the various data types. We apply empirical estimation of data covariance matrices and employ hierarchical scaling parameters to reduce the dependence on some assumptions required by the empirical approach. Hence, we account for noise variances and covariances, which is crucial for successful probabilistic multiphysics inversion. The parameter estimates and data covariance matrices are obtained with the reversible-jump Markov chain Monte Carlo algorithm with parallel tempering to enhance the efficiency. Since covariance matrix estimation changes data weights, the estimation process is carried out while samples are not recorded for inference.
The results at the Athabasca site fit the data and produce plausible data covariance matrices for the data weights. The structure of the lithosphere is resolved at various scales. The sedimentary basin is inferred with low uncertainties, constraining a thickness of 1.1 km. Discontinuities in shear-wave velocity are inferred at 8.3 km depth in the basement and at 35.5 km depth (the Moho). The lithosphere-asthenosphere boundary is inferred at 168 km depth as a resistivity contrast with no clear contrast in shear-wave velocity. A comparison to inversions that consider only seismic or magnetotelluric data shows reduced uncertainties on the positions and resistivities/velocities of all structures. Notably, combining magnetotelluric and surface-wave dispersion data reduces the uncertainty of the lithosphere depth, despite the absence of a shear-velocity contrast.

\end{summary}

%\begin{keywords}
% Magnetotellurics; Bayesian inference; Inverse theory; Joint inversion; Monte Carlo methods; Numerical modelling; Probability distributions; Statistical methods; Earthquake ground motions; Moho depth; Seismic discontinuities; Surface waves and free oscillations; Cratons
%\end{keywords}

%!!
%=======================================================================================================================
%!!
\section{Introduction}

Geophysical inverse methods vary in their ability to infer Earth's structure and processes. These variations arise from the different sensitivities of the various data types, and the assumptions made regarding the Earth model. For example, some seismic methods may infer structures based on impedance contrasts. However, inferring features such as the Lithosphere-Asthenosphere boundary (LAB) beneath cratons remains challenging since partial melting causes only relatively small impedance contrasts~\citep{eaton2009elusive}. On the contrary, magnetotelluric (MT) methods are highly sensitive to partial melting which reduces electrical resistivities, thereby providing a means to resolve the LAB. Therefore, joint inversion of various types of geophysical data is often applied to improve subsurface inferences by integrating their complementary information~\citep{gallardo2003characterization, linde2006improved,chen2012joint,afonso2016imaging}. Joint inversion can refer to the integration of data sets sensitive to the same physical parameters, such as different seismic and/or geodetic data types~\citep[e.g.,][]{julia2000joint,dettmer2016tsunami,vasyura2023bayesian}, or data sets governed by different physical processes~\citep[e.g.,][]{chen2012joint,linde2016joint}. For clarity, we refer to the latter as \emph{multiphysics inversion} (MPI) which is the subject of this paper.

The application of MPI requires care to effectively use the complementary information content of different data types with various resolutions without producing artifacts. An appropriate implementation includes flexible model parameterization~\citep{sambridge2006trans}, coupling of different physical processes and their parameters\citep{bodin2016imaging}, and objective relative weights for various data types. In this study, we apply trans-dimensional (trans-D) Bayesian inference to address these MPI challenges with the goal of integrating  magnetotelluric (MT), receiver function (RF), and surface wave dispersion (SWD) data to infer one dimensional (1D) models of lithospheric structure. Bayesian inference allows for robust uncertainty quantification when compared to linearized approaches for this strongly non-linear MPI problem.

An appropriate parameterization (e.g, the number of layers in a 1D model, the presence of anisotropy) that is consistent with data resolution is generally not known independently and inversion results generally have a strong dependence on the choice of parameterization and regularization~\citep{sambridge2006trans}. For the case of horizontally stratified earth models, trans-D models address this by treating the number of layers as unknown, thereby considering a solution to the inverse problem consisting of an ensemble of models with various dimensions. Inversion results exhibit natural parsimony and can be viewed as self- or data-regularized~\citep{malinverno2002parsimonious}. Even though trans-D approaches are widely applied in geophysics~\citep[e.g.,][]{agostinetti2010receiver, bodin2012transdimensional, dettmer2013transdimensional, rosas2014two, ghalenoei2021gravity}, applications of probabilistic MPI (PMPI) to field data are limited.

For one-dimensional Earth models, it is not always appropriate to assume a common parameterization for all parameter types. Assuming coincident 1D interfaces can result in full structural coupling between parameters. For example, while P-wave RF data can resolve the Moho well due to a sharp increase in seismic velocity, MT data may not be sensitive to the Moho due to a lack of electrical resistivity contrast across this interface. At the LAB, a reverse scenario is possible. P-wave RFs poorly resolve the LAB since the small velocity contrast does not generate strong conversions from P- to S-waves. However, MT data are highly sensitive to the presence of partial melt. Therefore, complete structural coupling can lead to overparameterized Earth models which can increase parameter uncertainties~\citep{dettmer2009analyzing}. Trans-D models can control the degree of structural coupling by permitting non-coincident (decoupled) interfaces, and hence separate spatial structures for different parameters consistent with data resolution and prior information. For example, \citet{bodin2016imaging} treat the number of layers and the presence of seismic anisotropy in each layer as unknown in the joint inversion of multiple seismic data types. \citet{piana2018flexible} employ a structural decoupling (SD) algorithm to infer the number of coupled interfaces and all possible classes of decoupled interfaces. They demonstrate the method for the case of two physical parameters (e.g, resistivity and shear-wave velocity) which result in three classes of interfaces, and provide hints to generalize the method to the case of more parameters. \citet{dettmer2014automated} nested two trans-D models to achieve data-driven decoupling with flexibility for any number and combinations of parameter types. In this study, we employ the latter approach.

Noise is an inseparable combination of measurement and theory errors. Theory errors arise from incomplete physics, discretization, and other model limitations. Robust estimation of noise levels is crucial for reliable probabilistic inference. An important challenge is that geophysical data, such as MT, RF, and SWD data can contain non-stationary correlated noise~\citep{egbert1997robust, bodin2012transdimensional, dettmer2012trans, dettmer2015direct}. Hierarchical parametric approaches have been used in geophysical Bayesian inversions by modeling the noise structure by a finite number of unknowns. The assumed noise model must represent the noise characteristics reasonably well while avoiding over-parameterizatione~\citep[e.g.,][]{bodin2012transdimensional, dettmer2012trans,ghalenoei2022joint}. In some cases, Monte Carlo modeling can be applied to characterize components of theory errors~\citep{duputel2014accounting}.

We present a new general method for PMPI that combines trans-D decoupling with noise estimation and automated data weighting. We follow~\citet{dettmer2014automated} by treating the number of layers and the types of parameters that change value across each interface as unknowns. The method is general, simple to implement, and can be used for any number of physical parameters. We invert for three physical parameters: electrical resistivity $\rho$, shear-wave velocity $V_s$, and the $V_p/V_s$ ratio, where $V_p$ is the compressional wave velocity. In addition, a nonparametric method is applied that combines empirical and hierarchical approaches to estimate the noise structure of data during inversion~\citep{dettmer2007uncertainty, vasyura2021accounting}. To address the dependence of the multiphysics inversion results on the relative weighting of the data sets~\citep{moorkamp2007joint}, we estimate data covariance matrices for the various data types. These matrices provide an objective weighting based on noise characteristics and are important for a successful PMPI.

The method was applied to MT, SWD and RF data to resolve the 1D lithospheric structure for both simulated data and field data from a region of the North American Craton in Alberta, Canada. The results demonstrate improvements in terms of inferring both the crustal structure and the LAB. The inversion results fit the data well with trans-D structural coupling and without the need for subjective data weights. The resistivities, seismic wave velocities and spatial extents of the main structures are well resolved with relatively low uncertainty. The main structures include the western Canadian Sedimentary Basin (WCSB), the crystalline basement, the Moho, and the LAB. Comparison to inversion results from individual data sets demonstrates significant advantages of PMPI.

%!!
%===============================================================================================================================
%!!
\section{Methods}
\subsection{Bayesian inference}
We employ Bayesian inference with a trans-dimensional model~\citep{sambridge2006trans} to quantify the geophysical properties of the subsurface. Bayesian inference is based on updating prior probabilities that are independent of the observed data with data information~\citep{jaynes2003probability}. In particular, we are interested in quantifying subsurface properties based on data from multiple physical processes, including seismic waves and electromagnetic signals. This process is often referred to as multiphysics inversion (MPI) and is generally nonlinear~\citep{vozoff1975joint, haber1997joint}. In this section, we demonstrate that Bayesian numerical methods are particularly suitable for nonlinear PMPI due to objective integration of the various data and flexible, data-driven structural coupling. We consider PMPI for the analysis of MT, RF, and SWD data. The unknown parameters of the subsurface include electrical resistivity $\rho$, shear-wave velocity $V_s$, and the $V_p/V_s$ ratio, where $V_p$ is the compressional wave velocity.

Consider $N$ observed data $\mathbf{d}$ and a group of model parametrizations $\mbf{m}_k$, with $k\in\mathcal{K}$ and $\mathcal{K}$ a countable set, where each model has $M_k$ parameters. In this case, Bayes' rule is~\citep{green1995reversible}
\begin{equation} \label{eq-bayes}
P(k,\mbf{m}_k|\mbf{d})=\frac{P(k)P(\mbf{m}_k|k)P(\mbf{d}|k,\mbf{m}_k)}{\sum_{k'\in\mathcal{K}}\int_\mathcal{M}P(k')P(\mbf{m'}_{k'}|k')P(\mbf{d}|k',\mbf{m}'_{k'})},
\end{equation}
where $P(k,\mathbf{m}_k|\mathbf{d})$ is the posterior probability density (PPD), $k$ indexes model choice with prior $P(k)$, $P(\mbf{k}_k|k)$ is the prior for geophysical parameters, and $P(\mathbf{d}|k,\mathbf{m}_k)$ is interpreted as the likelihood function $L(k,\mbf{m}_k)$. Since the PPD is high-dimensional and there is no closed-form solution for nonlinear problems, we employ reversible-jump Markov-chain Monte Carlo (rjMcMC) sampling with the Metropolis-Hastings-Green algorithm~\citep{brooks2011handbook}. 

\subsection{Likelihood function and covariance estimation for MPI} 
Under the assumption of independent errors between the various data types, the likelihood function for MPI can be written
\begin{equation} \label{eq2}
    L(\mathbf{m})=\prod_{i=1}^{D} l_i(k,\mathbf{m}_k),
\end{equation}
where $i$ indexes $D$ data sets. In this work, data include receiver function, SWD, and MT data. To formulate $l_i$, assumptions about the statistics of data errors are required. A reasonable assumptions for real-valued data are Gaussian distributed errors of the form
\begin{equation} \label{eq-like}
    l_i(k,\mathbf{m}_k)=\frac{1}{(2\pi)^{N_i/2}{|\mbf{C}_i|}^{1/2}}\exp\biggl [-1/2 \Bigl(\mathbf{d}_i-\mathbf{d}_i(\mathbf{m})\Bigr)^T\mbf{C}_i^{-1}\Bigl(\mathbf{d}_i-\mathbf{d}_i(\mathbf{m})\Bigr) \biggr],
\end{equation}
where $N_i$ are the number data for observed data $\mbf{d}_i$ and $\mathbf{d}_i(\mathbf{m})$ are data predictions. The $\mbf{C}_i$ are data covariance matrices that contain the statistical properties of data errors and play a crucial role in terms of the amount of information data contained. This is important for MPI since the $\mbf{C}_i$ determine how much each data type contributes to the solution $P(k,\mbf{m}_k|\mbf{d})$. We employ numerical estimation for the $\mbf{C}_i$ that are data based. Since data errors are not accessible independently, data residuals are considered a proxy and defined as $\mathbf{r}_i=\mathbf{d}_i-\mathbf{d}_i(\mathbf{m})$.

The MT data are considered in the frequency domain and are complex-valued. Under the assumption of Gaussian distributed errors in MT time-series, errors on complex impedance data in the frequency domain are circularly-symmetric Gaussian distributed errors, the likelihood is~\citep{dosso2011bayesian}
\begin{equation} \label{eq-Clike}
    l_i(k,\mbf{m}_k)=\frac{1}{\pi^{N_i}|\mbf{C}_i|}\exp\biggl [- \Bigl(\mbf{d}_i-\mbf{d}_i(\mbf{m})\Bigr)^\dagger \mbf{C}_i^{-1}\Bigl(\mbf{d}_i-\mbf{d}_i(\mbf{m})\Bigr) \biggr]
\end{equation}
where $\dagger$ denotes the conjugate transpose.

In this work, the $\mbf{C}_i$ (including variances and covariances) are estimated from the data to ensure data weights in PMPI that are not based on practitioner choice. Choosing weights subjectively is an important limitation in many joint inversions~\citep{moorkamp2007joint} and is avoided here. Estimating both variances and covariances is important since the RF, SWD, and MT data are known to include nonstationary and correlated errors due to a combination of measurement and theory errors~\citep[e.g.,][]{egbert1997robust, bodin2012transdimensional, dettmer2012trans, dettmer2015direct}.

We employ an automated empirical estimation of $\mbf{C}_i$ ~\citep{dettmer2007uncertainty} during sampling ~\citep{vasyura2021accounting}. The $\mbf{C}_i$ are periodically updated during the burn-in phase until convergence, based on only small changes occurring between updates. The algorithm initially assumes independent, identical errors and implicitly samples standard deviations for a pre-defined number of rjMcMC steps (20,000 in this work). Subsequently, updates to $\mbf{C}_i$ are computed for the highest-likelihood model for the most probable $k$. Nonstationary scaling parameters~\citep[eq.~(15)]{dettmer2007uncertainty} are applied based on a predefined window length (here: 1/5 of the number of data) and inversion results are not sensitive to this choice. Once the $\mbf{C}_i$ converged, the burn-in is considered complete and posterior sampling starts with fixed $\mbf{C}_i$. However, since the $\mbf{C}_i$ estimates are based on the maximum likelihood (ML) model for the peak in $k$, we apply hierarchical scaling parameters $s_i$ during sampling to reduce potential impacts due to these assumptions.

\subsection{Trans-dimensional approach and model parameterization} \label{sec-par}
PMPI presents challenges with respect to choosing an appropriate model parameterization. Trans-dimensional models are flexible and have been employed to treat the number of subsurface layers in 1D Earth models as unknown~\citep{malinverno2002parsimonious}. Recent works have considered trans-D sampling for structural decoupling~\citep{bodin2016imaging, piana2018flexible}. 

\begin{figure}
    \centering
    \includegraphics[width=1.0\linewidth]{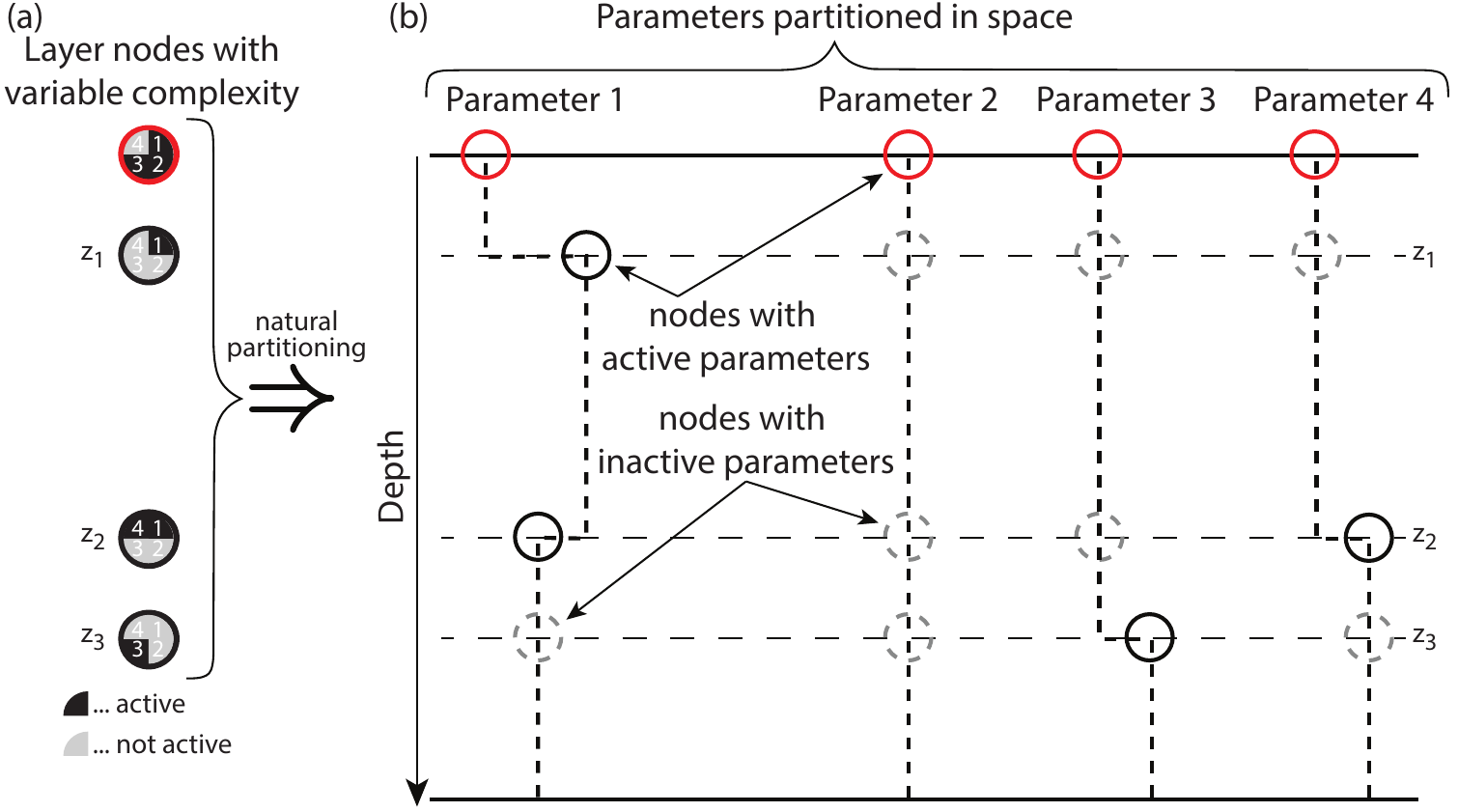}
    \caption{Schematic of variable complexity parametrization with trans-D layer nodes: (a) Layer nodes are specified for the depth of the top of the layer. The surface node (red) is always present. Additional parameters are activated (black) or deactivated (gray) by reversible-jump steps. Here, nodes include resistivity, shear-wave velocity, and Vp/Vs ratio. (b) The partitioning obtained from layer nodes. Layer nodes result in unique partitioning for each parameter type where discontinuities for that type are defined by the activated node parameters (solid circles). Values of the surface node define the half-space when no other parameters are active. Solid horizontal lines mark the limit of the partition and horizontal dashed lines indicate interfaces between layers, which are only discontinuities for a parameter type when a solid circle is shown.}
\label{fig-complexity}
\end{figure}

Here, we present an approach based on previous work~\citep{dettmer2014automated} where the layer nodes define the subsurface model (Fig.~\ref{fig-complexity}), and refer to this approach as variable complexity sampling. Each node defines the top of a subsurface layer and may contain active and inactive parameters. In our case, the node parameters are depth, electrical resistivity $\rho$, shear-wave velocity $V_s$, and the $V_p/V_s$ ratio, where $V_p$ is the compressional wave velocity. The parameters on the nodes are activated or deactivated by reversible-jump steps, and the number of nodes is changed by reversible jump steps (details in Supplement A). We employ proposal distributions for reversible-jump steps that are equal to the prior distributions. Both the number of layer nodes and the types of active parameters are sampled with the same algorithm without the need for additional considerations (Fig.~\ref{fig-complexity}). The concept is simple yet general and applies to parameters that describe the subsurface continuum, discrete parameters (e.g., dip angles are not considered in this work). The concept extends to two or three spatial dimensions by employing Voronoi nodes~\citep{sambridge1998tomographic}. In this work, the surface node is always present and defines the half-space in the absence of other nodes. Furthermore, depth profiles of parameter types of interest are defined only by the nodes where that type is active. Therefore, the method provides decoupled partitioning based on the data information. 

We consider the case of one-dimensional (1D) multiphysics inversion of MT, SWD, and RF data at lithospheric scales with three parameter types ($\rho$, $V_s$, and $V_p/V_s$) and three hierarchical noise parameters. In this case, MT data are typically not sensitive to the location of the Moho but are sensitive to the location of the LAB. In contrast, RF data are sensitive to the Moho but often relatively insensitive to the LAB. Such differences are typical for MPI~\citep{moorkamp2017integrating}. Therefore, applications of probabilistic MPI may benefit from the data-driven decoupling we propose here. Such issues are not limited to MPI. Individual data types generally exhibit various degrees of resolution for various parameter types. For example, the SWD data resolve $V_s$ well, but $V_p/V_s$ poorly. Structural decoupling can result in more appropriate uncertainty quantification by reflecting these differences. However, activating all layer-node parameters results in full structural coupling, which may be desirable in some cases.
%!!
%========================================================================================
%!!

\subsection{Forward models}
Every rjMcMC step requires predicting data for the proposed model. The MT impedance for each frequency was computed using the analytic recursive formula obtained by solving Maxwell's equations with appropriate boundary conditions on a stack of homogeneous 1D layers~\citep[e.g., Equation~5.25]{zhdanov2009geophysical}. For SWD data, we computed the fundamental mode Rayleigh wave phase velocities, based on calculating the normal modes of the Earth~\citep{saito1988disper80}. We considered SWD periods up to 250 s. Since these periods are sensitive to the structure significantly deeper than the maximum depth of interest, we appended the preliminary reference Earth model (PREM)~\citep{dziewonski1981preliminary} below 250-km depth to depths of 1000 km~\citep{dettmer2015direct}. For RF computations, we first calculated horizontal and vertical seismograms with the ray-based method of~\citet{frederiksen2000modelling}, which has the flexibility of including only phases with significant amplitudes. We included the direct P- to S-wave conversion phase and multiples that contain significant energy~\citep{dettmer2015direct}. After computing seismograms, the vertical component was deconvolved from the radial component by water-level deconvolution~\citep{langston1979structure}. For seismic forward calculations, density was constrained by Birch's law~\citep{birch1960velocity}.

\section{Results} \label{sec_results}
This section considers the inversion of simulated and field data. To improve rjMcMC convergence, we employed parallel tempering (PT) ~\citep{dettmer2012MFI, sambridge2014parallel}, which uses a series of parallel chains, whose target likelihoods are raised to powers of $0<\beta\leq1$, and swaps between chains are allowed to improve exploration. By decreasing $\beta$, chains favour wider sampling, which improves sampling at the target distribution ($\beta=1$). For simulations and field data, we employed 12 chains in parallel, five of which were at $\beta=1$, and $\beta$ was sequentially decreased by a factor of 0.85 for six tempered chains. A master process collected posterior samples from chains at $\beta=1$. The acceptance ratios for all proposal types during the inversion were monitored and ratios around 35-40$\%$ were achieved. Convergence was judged by monitoring the recorded likelihood values and the number of populated nodes for each parameter. After discarding the non-equilibrium burn-in phase (at least 500,000 samples for field data), the distribution of the first and the last 1/3 of samples were approximately the same~\citep{dettmer2013transdimensional}.

%!!
%=======================================================================================
%!! 
\subsection{Simulated data} \label{sec-sims}
\begin{figure}
    \centering
    \includegraphics[width=\linewidth]{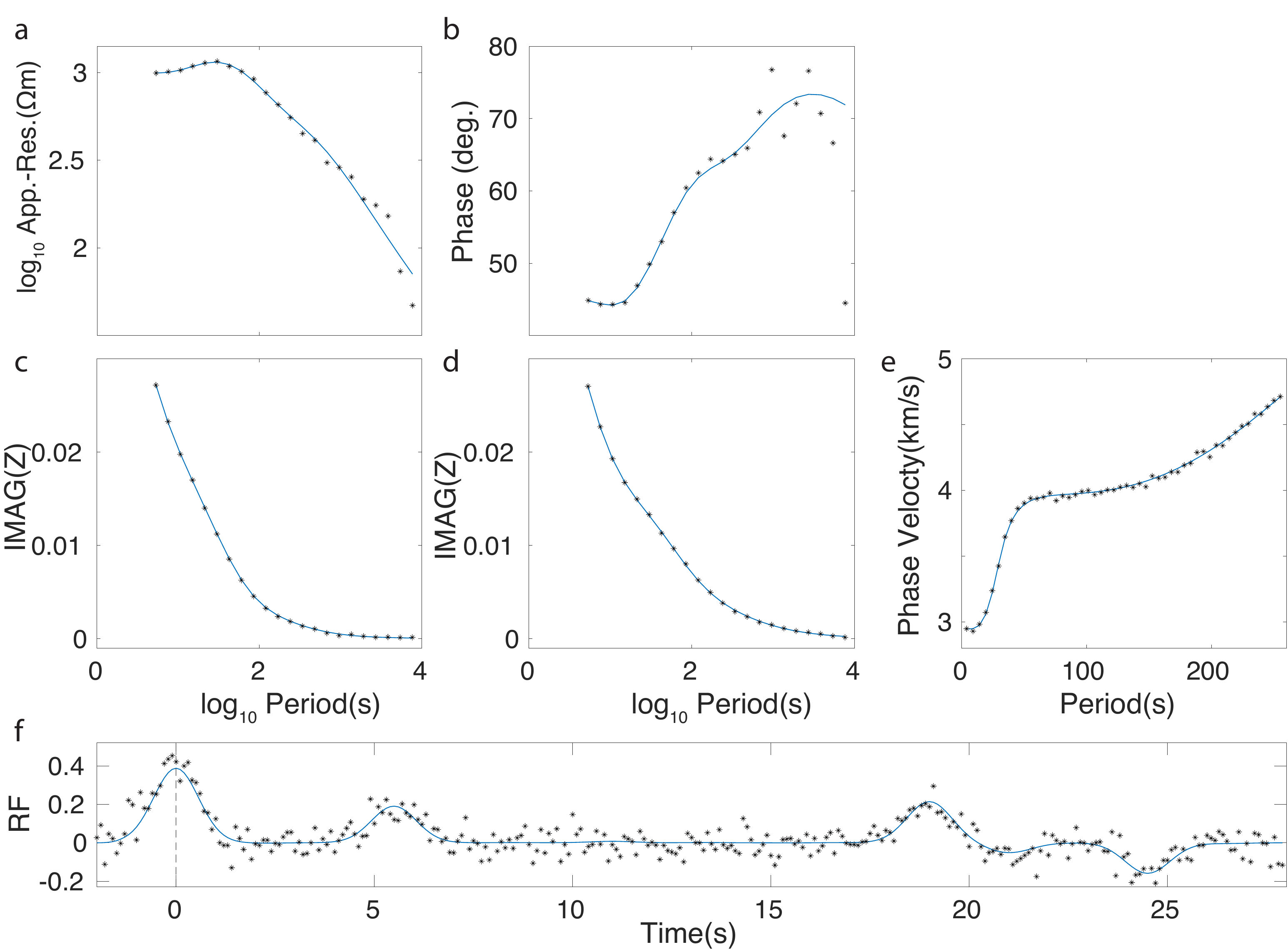}
    \caption{The true (blue lines) and noisy (black dots) data for the simulation study. (a) Apparent resistivities and (b) phases of MT data. MT data used in the inversions are complex impedance values shown as (c) real and (d) imaginary parts. (e) SWD and (f) RF data are also shown.}
    \label{fig-SIM-data}
\end{figure}

\begin{figure} 
    \centering
    \includegraphics[width=\linewidth]{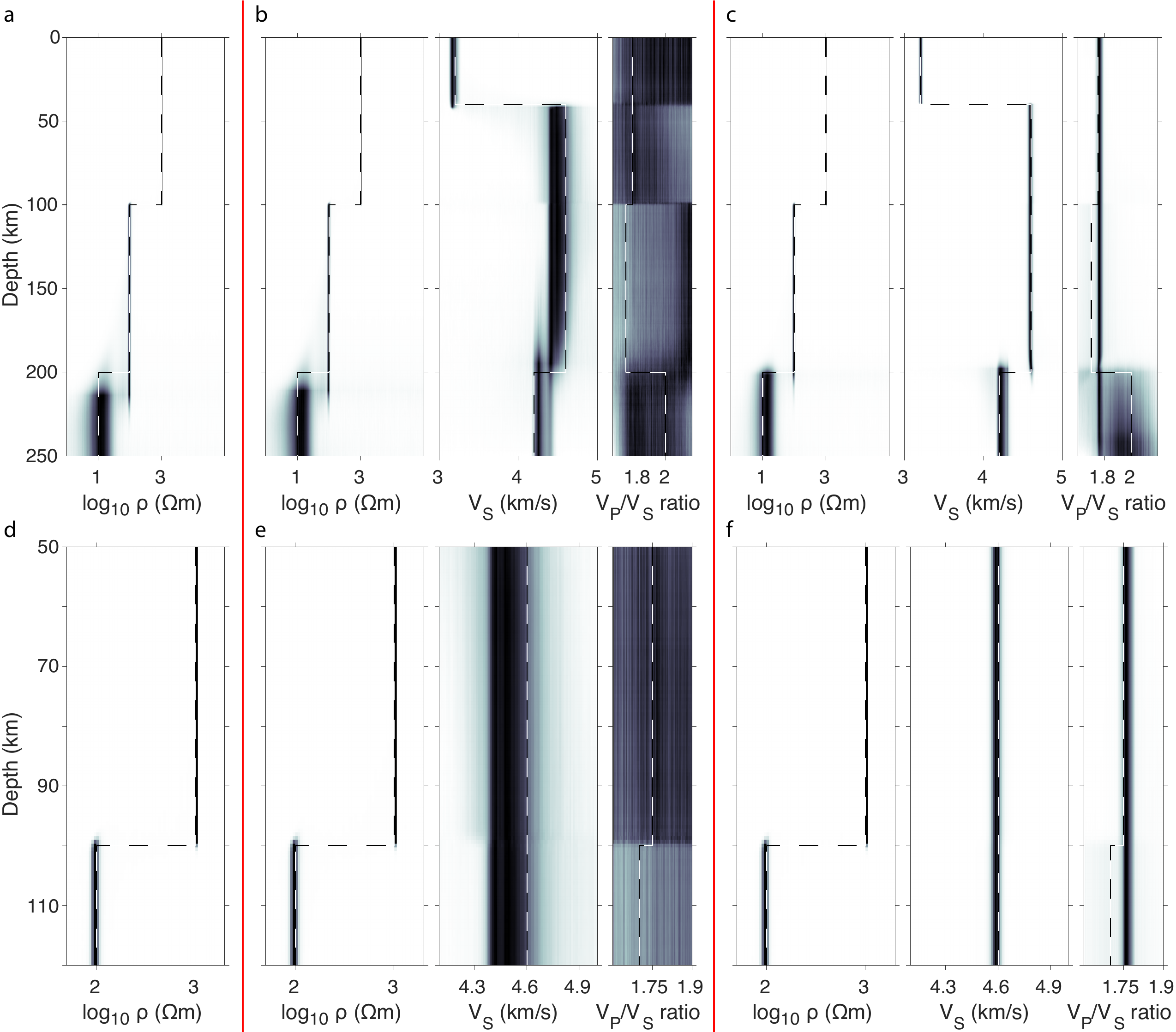}
    \caption{Posterior marginal profiles for logarithms of resistivities $\rho$, shear-wave velocities $V_s$, and $V_p/V_s$ rations from various combinations of simulated data: (a, d) Inversion of only MT data, (b, e) inversion of MT and SWD data, and (c, f) inversion of MT, SWD, and RF data. Panels a-c show the full depth range and panels d-f magnify the 50- to 120-km depth. Prior bounds are given by the axes extents in a-c and the true model is shown (dashed line).}
    \label{fig-SIM-marg}
\end{figure}

First, we demonstrate the algorithm with simulated data (Fig.~\ref{fig-SIM-data}) for a model where $\rho$ and $V_p/V_s$ have discontinuities that are different to those of $V_s$ (Fig.~\ref{fig-SIM-marg}). At 40-km depth only $V_s$ changes, at 100-km depth $\rho$ and $V_p/V_s$ change, and at 200-km depth all three parameters change. Data were contaminated by random Gaussian noise with standard deviations of $10^{-4}$ V/mT for MT impedances, 0.02 km/s for SWD phase velocities and 0.05 s$^{-1}$ for RF data (Fig.~\ref{fig-SIM-data}). Noise was assumed to be independent, identically distributed during the inversion and the standard deviations $s_i$ were treated as unknown parameters. We used a truncated Poisson distribution~\citep{green1995reversible} with scale parameter 3 in the interval [2, 30] for the prior on the number of layer nodes and a discrete uniform distribution on [1,3] for the number of active parameters within each node. Uniform priors were used for all other parameters (Table~\ref{table-priors}). To allow for a more flexible specification of priors for trans-D models, $V_s$ and $V_p/V_s$ were parameterized in terms of perturbations from an assumed background model (i.e., $dV_s$ and $dV_p/V_s$). However, inversion results do not depend on the choice of background model~\citep{dettmer2015direct}.

\begin{table}
    \centering
    \caption{Uniform priors used for all inversions. Note that the $s_\cdot$ are blank for field data since we estimate full data covariance matrices in that case.}
    \begin{tabular}{  c | c | c  }
           
        Quantity & Simulated data & Field data\\
        \hline
        $\log_{10}\rho\ (\Omega m)$ & [0.0, 5.0]  & [0.0, 3.7]\\
        $dV_s\ (km/s)$ & [-1.0, 1.0] & [-0.1, 0.1]\\
        $dV_p/V_s$ & [-0.3, 0.3] & [-0.1, 0.1]\\ 
        $s_{MT}\ (V/(mT))$ & $[10^{-5}, 10^{-3}]$ & -\\
        $s_{SWD}\ (km/s)$ & [0.001, 0.1] & - \\ 
        $s_{RF}\ (s^{-1})$ & [0.001, 0.1] & -\\
    \end{tabular}
    \label{table-priors}
\end{table}

\begin{figure}
    \centering
    \includegraphics[width=\linewidth]{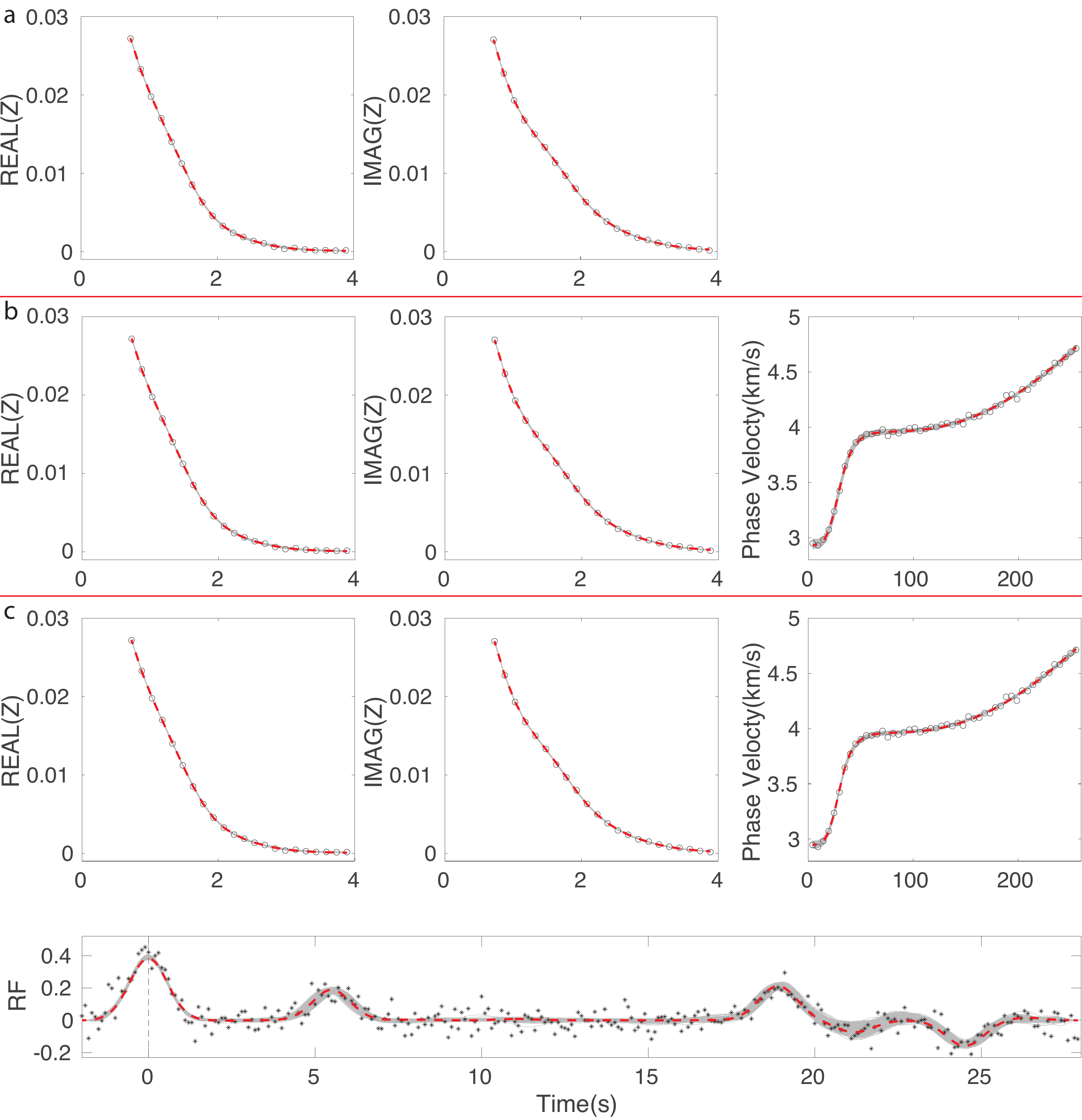}
    \caption{Simulated data and posterior predictive distributions for (a) only MT data, (b) MT and SWD data, and (c) MT, SWD, and RF data. Predictions are shown in gray for all data types. The best fitting models, i.e., the models with the highest likelihood for the most probable number of nodes, are shown with red dashed lines.}
    \label{fig-SIM-postpred}
\end{figure}

\begin{figure}
    \centering
    \includegraphics[width=\linewidth]{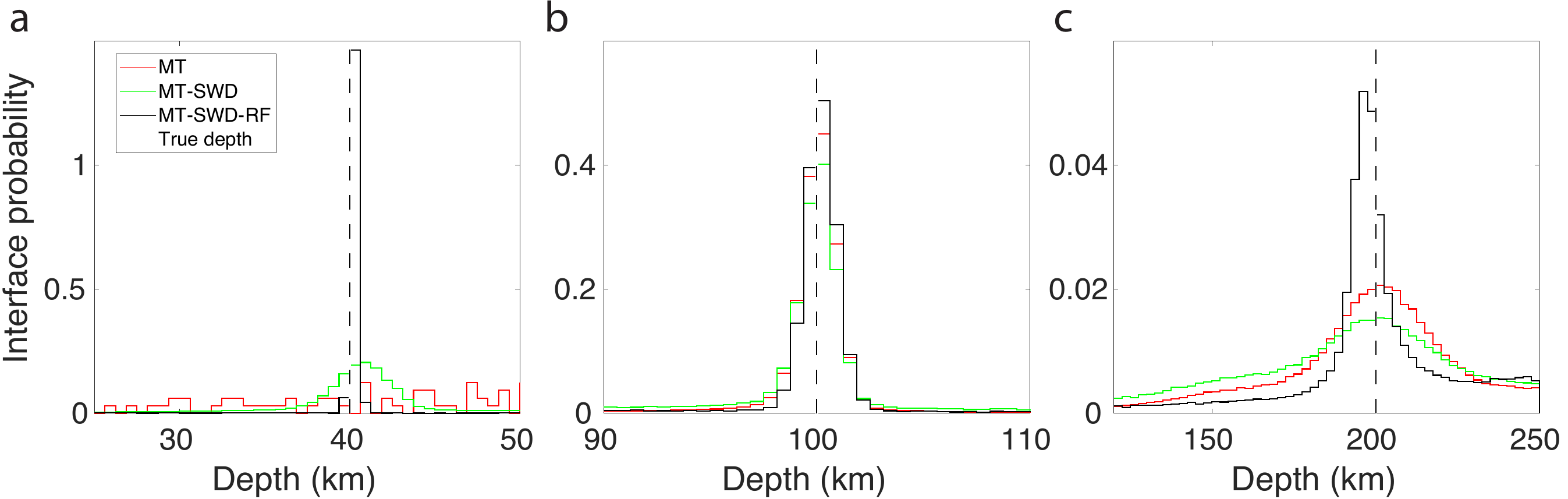}
    \caption{Interface probabilities for depth ranges from (a) 25--50 km, (b) 90--110 km, and (c) 130--250 km for simulated data. Results for MT only (red), MT-SWD (yellow), and MT-SWD-RF (black) are shown. All distributions in each panel are normalized to unit area to enable comparison of uncertainties. True values are shown by dashed black lines.}
    \label{fig-SIM-iface}
\end{figure}

\begin{figure}
    \centering
    \includegraphics[width=\linewidth]{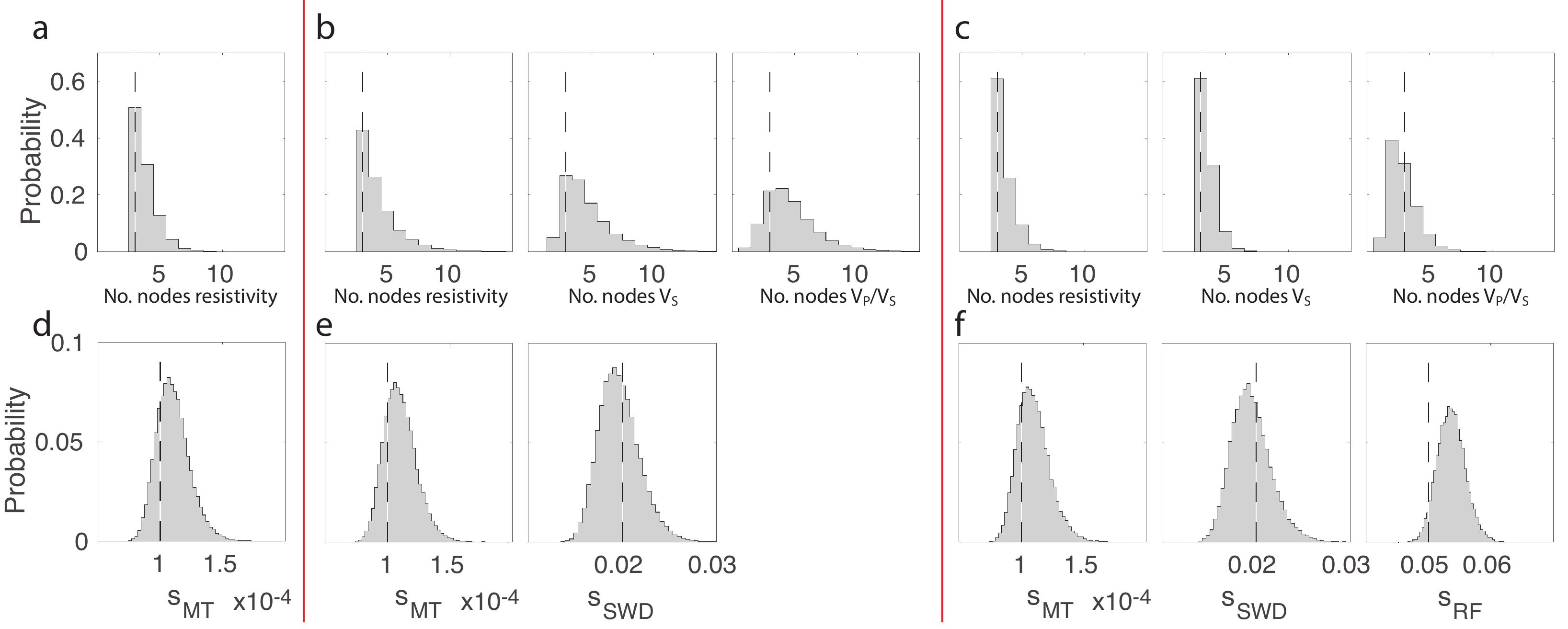}
    \caption{Marginal distributions for (a--c) the number of nodes per parameter type and (d--f) standard deviation parameters for simulated data. Inversion results for (a, d) only MT data, (b, e) SWD and MT data, and (c, f) RF, SWD, and MT data. True parameter values are shown by dashed black lines.}
    \label{fig-SIM-k_sig}
\end{figure}

Since the focus of this work is on PMPI, we consider three cases in the main text: (1) only MT data, (2) MT and SWD data, and (3) MT, SWD and RF data. The results are described based on several figures (Figs~\ref{fig-SIM-marg} to~\ref{fig-SIM-k_sig}) that summarize the three cases. Supplement B contains the cases of only SWD data, only RF data, and SWD and RF data for reference (Figs S1 to S4).

Figure~\ref{fig-SIM-marg} shows the marginal probability profiles for resistivity, shear-wave velocity, and $V_p/V_s$ ratio as a function of depth. Resistivity and shear-wave velocity profiles for all cases show three layers with discontinuities at $\sim$40 km for $V_s$, $\sim$100 km for $\rho$, and at $\sim$200 km for all parameters, which agrees with the true partitioning (Figs~\ref{fig-SIM-marg}a to c). Compared to other cases, Case 3 significantly decreases the uncertainty for $V_s$ and reduces the resistivity uncertainty, particularly for depths greater than $\sim$200 km. The $V_p/V_s$ uncertainty in case 2 is high, covering the prior width for all depths. In general, only case 3 (Figs~\ref{fig-SIM-marg}c and f) significantly decreases the uncertainty of $V_p/V_S$. However, it cannot resolve the 100-km discontinuity from 1.75 to 1.7, and the amount of uncertainty at 200-km depth is large, demonstrating low sensitivity to $V_p$. The most probable profile for $V_p/V_s$ is a half-space with a value of $\sim$1.75. Case 2 (Figs~\ref{fig-SIM-marg}b and e) shows evidence of an erroneous $V_s$ discontinuity at 100-km depth. Case 3 removes this effect (Fig.~\ref{fig-SIM-marg}f). In summary, Fig.~\ref{fig-SIM-marg} shows that complementary information in MT, SWD, and RF data reduces uncertainty in PMPI. This conclusion also holds when comparing the results with the cases of inversions of only SWD, only RF, and SWD-RF (Fig. S1).

Figure~\ref{fig-SIM-postpred} shows the data fits in terms of posterior predictive distributions that quantify the predictive power of the posterior ensemble and illustrate that data are fit to the noise level. This figure provides further evidence that PMPI produces comparable MT data fits compared to case 1 (Fig.~\ref{fig-SIM-postpred}a), and case 3 fits SWD data to a level similar to case 2. Similar observations are also made for the cases that only consider seismic data (Fig. S2).

To consider how PMPI affects uncertainties at discontinuities, Fig.~\ref{fig-SIM-iface} shows the marginal probabilities of the depths of the discontinuity. Unlike RF data, which have relatively high sensitivity to $V_s$ discontinuities (Fig. S3), MT and SWD data are of an integrative nature. Hence, including RF data improves interfaces resolution, particularly at 40-km and 200-km depths (Figs~\ref{fig-SIM-iface}a and c). Figure~\ref{fig-SIM-iface}b shows that case 3 does not reduce the uncertainty of the interface at 100-km depth across which only resistivity and $V_p/V_s$ change. This is consistent with the fact that the RF and SWD data are largely insensitive to compressional waves and therefore the $V_p/V_s$ ratio.

The marginal probability distributions of the number of nodes populated for each type of parameter (Figs~\ref{fig-SIM-k_sig}a to c) show that case 3 exhibits the lowest uncertainty on the number of nodes for all three parameters. Figures~\ref{fig-SIM-k_sig}d to f show marginal probabilities for data standard deviations, which differ significantly from their priors, indicating that the data contain information about noise parameters. Notably, the PMPI of case 3 (Fig.~\ref{fig-SIM-k_sig}f) does not diminish the fit to the data. 

We have included an additional simulation study in the supplement (Figs S5 to S10) that closely mimics the structure that was inferred at Athabasca, including frequency bandwidth and a sedimentary basin layer. The results for the second simulation are broadly consistent with the first simulation and confirm that the data can resolve a 1.5-km thick sedimentary basin structure as well as a Moho and a LAB. 

\subsection{Lithospheric structure near Athabasca, Canada} \label{sec-realdata}
\begin{figure}
    \centering
    \includegraphics[width=1.0\linewidth]{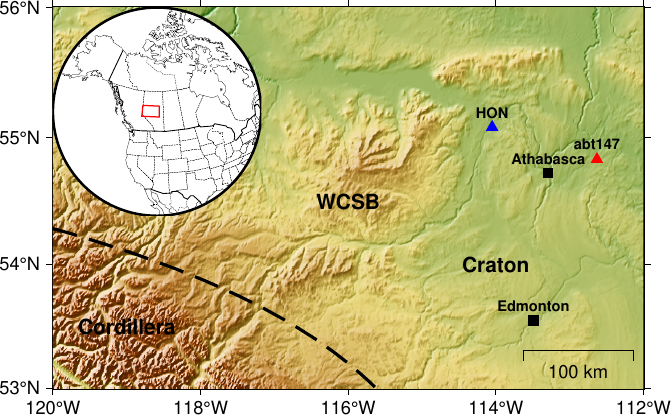}
    \caption{Map of the study location on the North American Craton near Athabasca, Alberta, Canada. The MT data were observed at the triangle marker labeled abt147. The SWD data were obtained for the area in the vicinity of the two triangles. The RF data were obtained from station HON. The dashed curve is the approximate boundary between the North American Craton and Cordillera.}
\label{fig-map}
\end{figure}

We consider field data for the three data types recorded near Athabasca City (Alberta, Canada) on the North American Craton that includes sedimentary rocks of the Western Canada Sedimentary Basin
(Fig.~\ref{fig-map}).
The MT data were recorded by Lithoprobe~\citep{boerner1999electrical,boerner2000synthesis} and the University of Alberta~\citep{wang2022three} at 54.83$^\circ$ N and 112.63$^\circ$ W over 22 periods in the range 5-7,680 s. Their dimensionality analysis~\citep{wang2022three} shows that they are consistent with 1D resistivity structure. Surface wave dispersion data were taken from the continent-scale velocity model SL2013NA~\citep{schaeffer2014imaging}. Fundamental mode Rayleigh wave phase velocities were calculated for the velocity model beneath the location of the MT data for 50 periods between 4 s and 255 s. For RF data, we chose one of the closest stations, HON, from the Y5 network at 55.08$^\circ$ N and 114.05$^\circ$ W~\citep{gu2011canadian}. This is the closest station to the MT data that provides high-quality waveforms~\citep{chen2015crustal}. We analyzed 951 teleseismic earthquakes with moment magnitude $M_w>5.5$ between 2007 and 2013 for epicentral distances between 30$^\circ$ and 90$^\circ$. We obtained P-wave receiver functions using the method of water-level deconvolution and employing a Gaussian filter with the width parameter 1.25~\citep{ammon1990nonuniqueness}. After manual quality control, 87 receiver functions with ray parameters between 0.045s/km and 0.075s/km were stacked to obtain the final receiver function.

As previously, we consider three cases: (1) only MT data, (2) MT and SWD data, and (3) MT, SWD and RF data. We assumed a truncated Poisson distribution with scale parameter 3 in the interval [2, 30] for the prior on the number of layer nodes for each parameter type. Uniform priors were assumed for all other parameters (Table~\ref{table-priors}). 
We chose independent identically distributed errors with realistic standard deviation values for simulations in the previous section to clearly illustrate the concept of PMPI with variable layer complexity. However, independent identically distributed errors are not realistic for field data. Therefore, full covariance matrices $C_i$ were estimated iteratively during the burn-in phase of rjMcMC, and hierarchical scaling parameters $s_i$ were sampled using uniform priors on [1., 10.] for all datasets. That is, covariances were estimated empirically as part of the sampling but a hierarchical scaling was introduced to reduce the dependence on the initial assumption of independence.

\begin{figure}
    \centering
    \includegraphics[width=\linewidth]{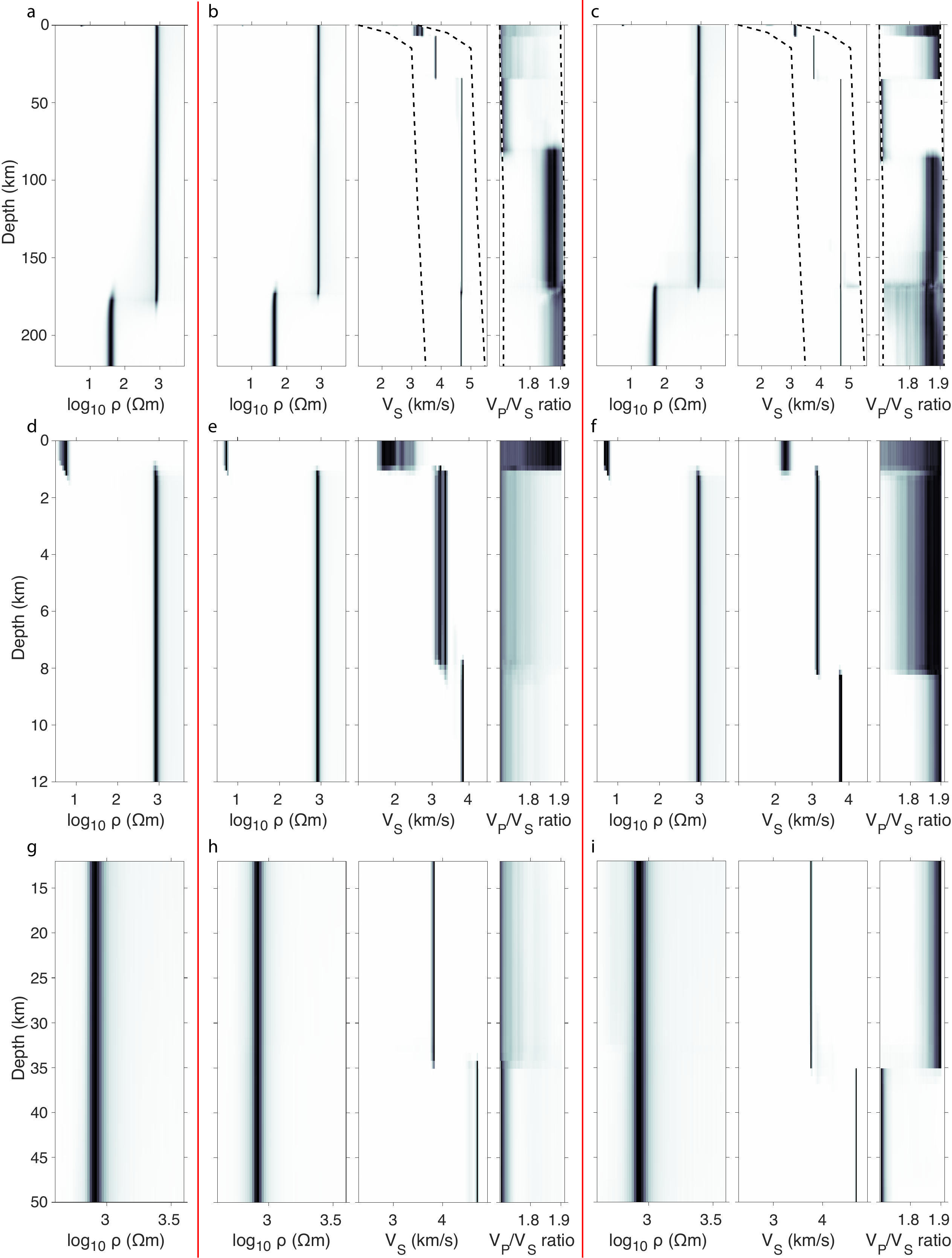}
    \caption{Posterior marginal distributions for a location near Athabasca, Canada for the (a--c) full depth considered in the inversion, (d--f) magnified for the uppermost 12 km, and (g--i) magnified for depths from 12 to 50 km. Results are presented for (a, d, g) only MT data, (b, e, h) MT and SWD data, and (c, f, i) MT, SWD, and RF data. Priors bounds on $V_s$ and $V_P/V_s$ are not given by plot bounds but shown by dashed black lines.}
    \label{fig-atha-marg}
\end{figure}

\begin{figure}
    \centering
    \includegraphics[width=\linewidth]{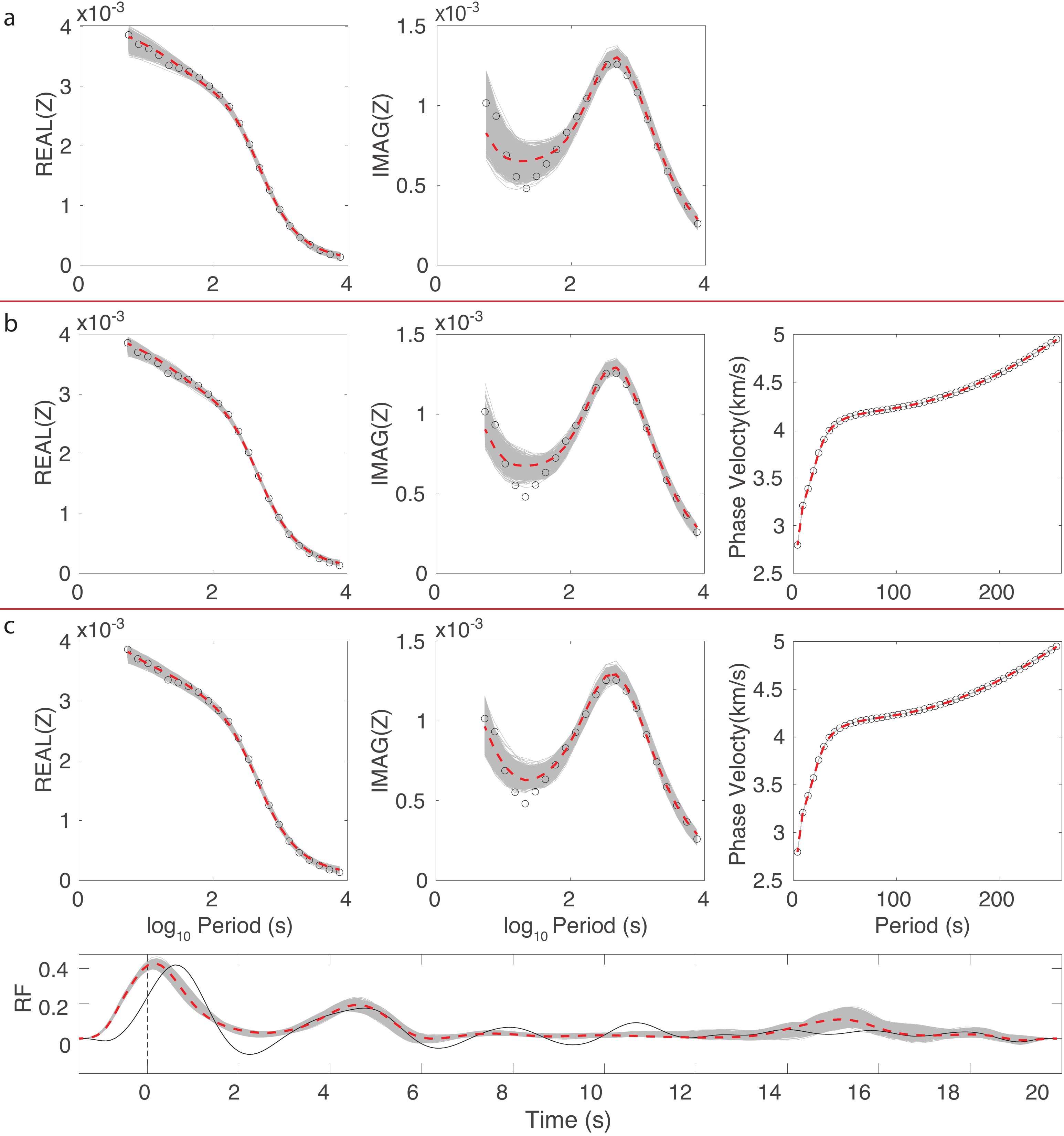}
    \caption{Athabasca data and posterior predictive distributions for (a) only MT, (b) MT-SWD, and (c) MT-SWD-RF. Field data for MT and SWD are shown by open circles while the RF is shown as a solid line. Predictions are shown in gray for all data types. The best fitting models, i.e., the models with the highest likelihood for the most probable number of nodes, are shown with red dashed lines.}
    \label{fig-atha-postpred}
\end{figure}

\begin{figure}
    \centering
    \includegraphics[width=\linewidth]{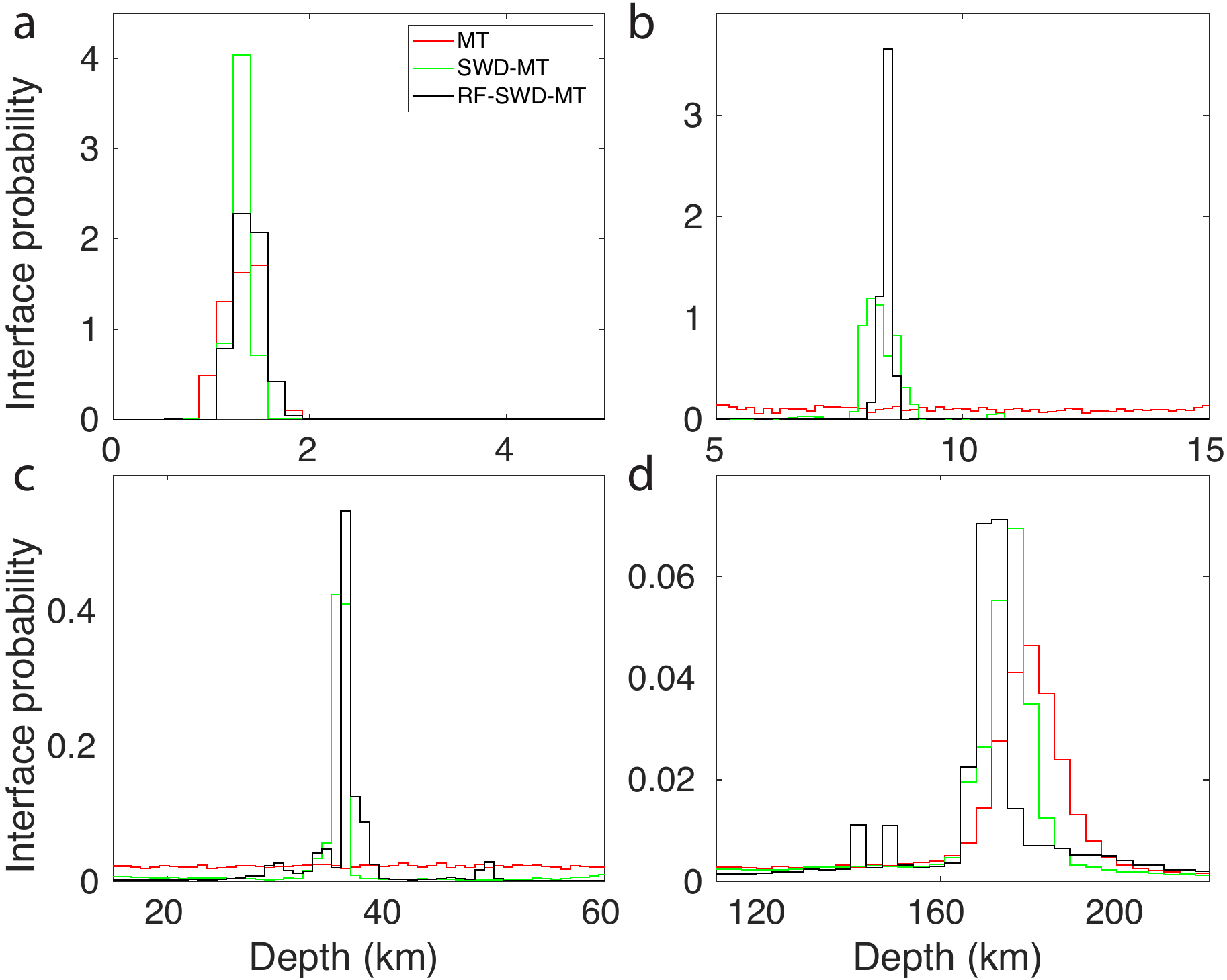}
    \caption{Interface probabilities for depth ranges from (a) 0--5 km, (b) 5--15 km, (c) 15--60 km, and (d) 110--220 km for data near Athabasca. Results for MT only (red), MT-SWD (yellow), and MT-SWD-RF (black) are shown. The distributions in each panel are normalized to unit area to enable comparison of uncertainties.}
    \label{fig-atha-iface}
\end{figure}

In case 1,  MT data constrain two resistivity discontinuities, including a shallow discontinuity near $\sim$1.2-km depth and a deep discontinuity at $\sim$176-km depth (Figs~\ref{fig-atha-marg}a, d, and g). The resistivity values of 3.9-7.1 $\Omega m$ in the uppermost layer are consistent with sedimentary basin values. Furthermore, the depth of this discontinuity is consistent with the depth of the WCSB in this area. We interpret the second discontinuity as the LAB since a clear reduction of resistivity occurs. 
Figure~\ref{fig-atha-postpred} shows posterior predictive distributions for the Athabasca data and all cases. Overall, good agreement is achieved between the field observations and the range of predictions produced by the PMPIs. However, a shift in the P-wave arrival for the RF data is not fit by the predictions. This figure provides further evidence that PMPIs produce comparable MT data fits when compared to case 1. Furthermore, case 3 fits SWD data to a similar level as observed in case 2.
Figure~\ref{fig-atha-iface} shows interface probability as a function of depth. For case 1, the WCSB discontinuity has uncertainty between 0.7 km and 1.75 km and the LAB discontinuity has uncertainty between 150 km and 205 km. 

For case 2 (Figs~\ref{fig-atha-marg}b, e, and h), the number of resistivity discontinuities is the same as case 1, but resistivity has lower uncertainty in the WCSB, between  4.7 and 6.0 $\Omega$m. The $V_s$ profile shows three sharp discontinuities which do not all coincide with the resistivity discontinuities. The $V_s$ of the first layer is resolved with high uncertainty between 1.5 km/s to 2.6 km/s, which is expected due to low sensitivity of the SWD data to the shallowest structures. However the low velocity values are consistent with a sedimentary basin and inferring this layer by PMPI is remarkable in itself as the inversion of only SWD data does not resolve it(Fig. S11k). The second and third layers have velocity uncertainties between 3 to 3.45 km/s and 3.72 to 3.85 km/s, respectively. The discontinuities for resistivity and shear-wave velocity closely agree at 1.1-km depth. Figure~\ref{fig-atha-iface} shows that thickness of the WSCB has decreased uncertainty between 0.88 and 1.4 km compared to case 1. The second $V_s$ interface has uncertainty between 7.7 and 9 km (Fig.~\ref{fig-atha-iface}). The third $V_s$ interface has uncertainty between 32 and 37 km. At a depth of 155-195 km, the $V_s$ profile does not show a clear discontinuity but $\rho$ exhibits a clear change (Figs~\ref{fig-atha-marg} and~\ref{fig-atha-iface}). Based on these values, we interpret the second discontinuity at $\sim$8-km depth as a change in crystalline basement rocks. The sharp velocity increase from 3.85 to 4.75 km/s at $\sim$37-km depth is the Moho.

For case 3, a noticeable reduction in uncertainty occurs (Figs~\ref{fig-atha-marg}c, f, and i) for $V_s$ values, particularly for the WCSB (2.1-2.47 km/s) and the upper crystalline basement rocks (3-3.25 km/s). Figure~\ref{fig-atha-iface} shows that the addition of RF data has reduced uncertainty on the depth of the WCSB compared to case 1 but has caused slightly higher uncertainty compared to case 2. This is likely a trad-off with the lower $V_s$ uncertainties in case 3 for this layer. The crystalline basement discontinuity is estimated with low uncertainty near 8.3 km. The depth of the Moho is inferred by case 3 at 35.5-km depth with uncertainty between 32 km and 39 km. The LAB is inferred similarly to case 2 but with the most probable depth of 168 km. The inversion of only SWD data, shows no evidence of the WCSB and LAB, and produces implausibly low uncertainties (Figs S11e, k, and q). The poor uncertainty estimates for the inversion of only SWD data are due to the fact that the data are derived from a velocity model, thereby they do not contain noise. Covariance estimation alleviates these issues to some degree but not completely. Inversion of only RF data (Figs S11d, j, and p) demonstrates high sensitivity to discontinuities but limitations to constrain a plausible $V_s$ profile. Inversion of RF and SWD data (Figs S11f, l, and r) shows some of the advantages of PMPI case 3 but lacks clear evidence for the LAB and shows higher uncertainty for the other structures (Fig. S13). 

For reference, the regional SL2013NA velocity model used to derive the SWD data is shown in Figure S15 in the supplement. The discrepancies between the inversion results and the SL2013NA model are due to the sensitivity of the SWD data.  The largest discrepancy between the SL2013NA $V_s$ model and the inversion results is for the upper 8 km. It can be explained by the low sensitivity of SWD data to shallow structures. The SL2013NA $V_s$ model shows a gradual change from 3.85 km/s to 4.75 km/s in the Moho by first increasing to 4.1 km/s at 32 km depth and then increase to 4.75 km/s at 41 km depth. $V_s$ uncertainties for only SWD and SWD-RF inversion cases (Figs S13n and o) are consistent with this gradual change. MT-SWD and MT-SWD-RF (Figs S13k and l) show a sharp increase in Moho, however, the uncertainty between 32 km and 39 km obtained for Moho (Figs~\ref{fig-atha-iface}c and S13c) is consistent with the SL2013NA velocity model.

\begin{figure}
    \centering
    \includegraphics[width=\linewidth]{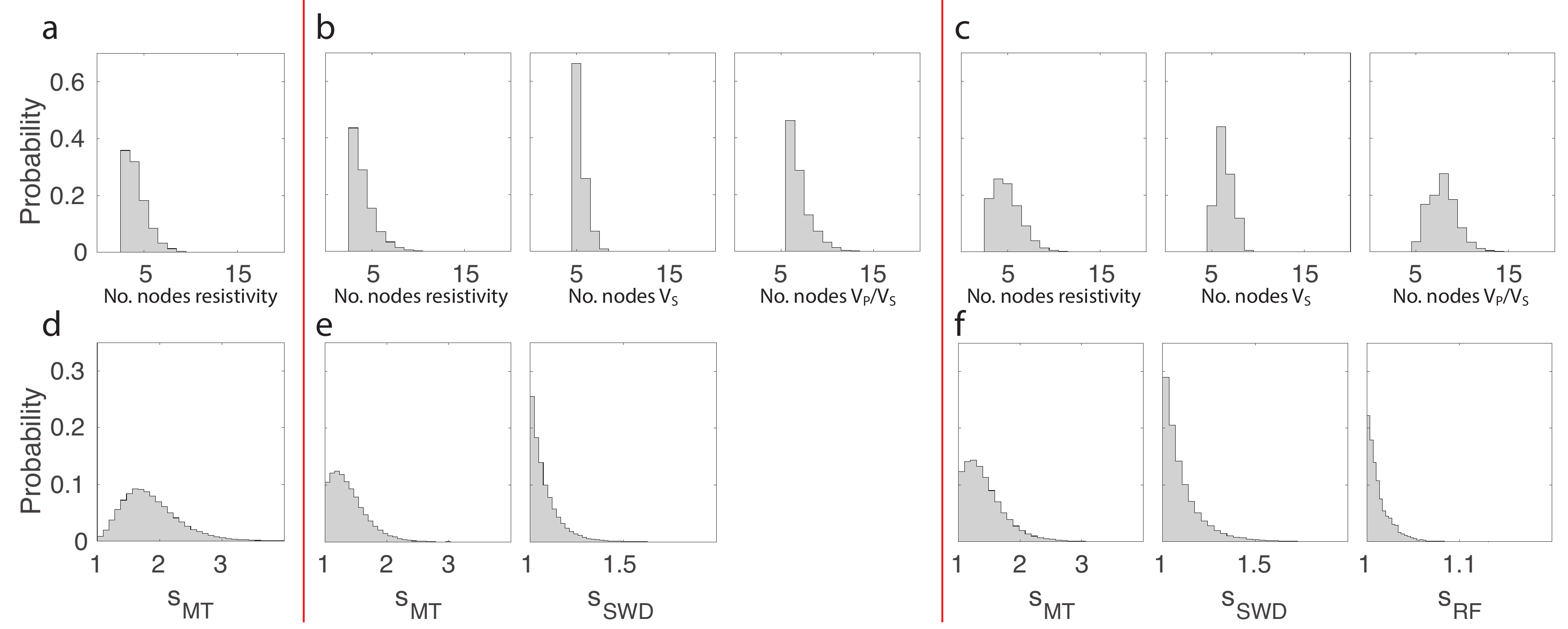}
    \caption{Marginal distributions for (a--c) the number of nodes per parameter type and (d--f) standard deviation parameters for Athabasca data. Inversion results for (a, d) only MT data, (b, e) SWD and MT data, and (c, f) RF, SWD, and MT data.}
    \label{fig-atha-k-sig}
\end{figure}

Figure~\ref{fig-atha-k-sig} shows the marginal probability distributions of the complexity of the model for each parameter. For cases 1 and 2, the most probable partitioning for resistivity has three nodes. In case 2, five and six nodes are the most likely for $V_s$ and $V_p/V_s$, respectively. In case 3, the uncertainty for the nodes increases, suggesting some inconsistencies between data types. Marginal probability densities for scaling factors of the data covariance matrices estimated as part of the PMPI (Figs~\ref{fig-atha-k-sig}d to f) decrease slightly from case 1 to case 2 and case 3.

\begin{figure}
    \centering
    \includegraphics[width=\linewidth]{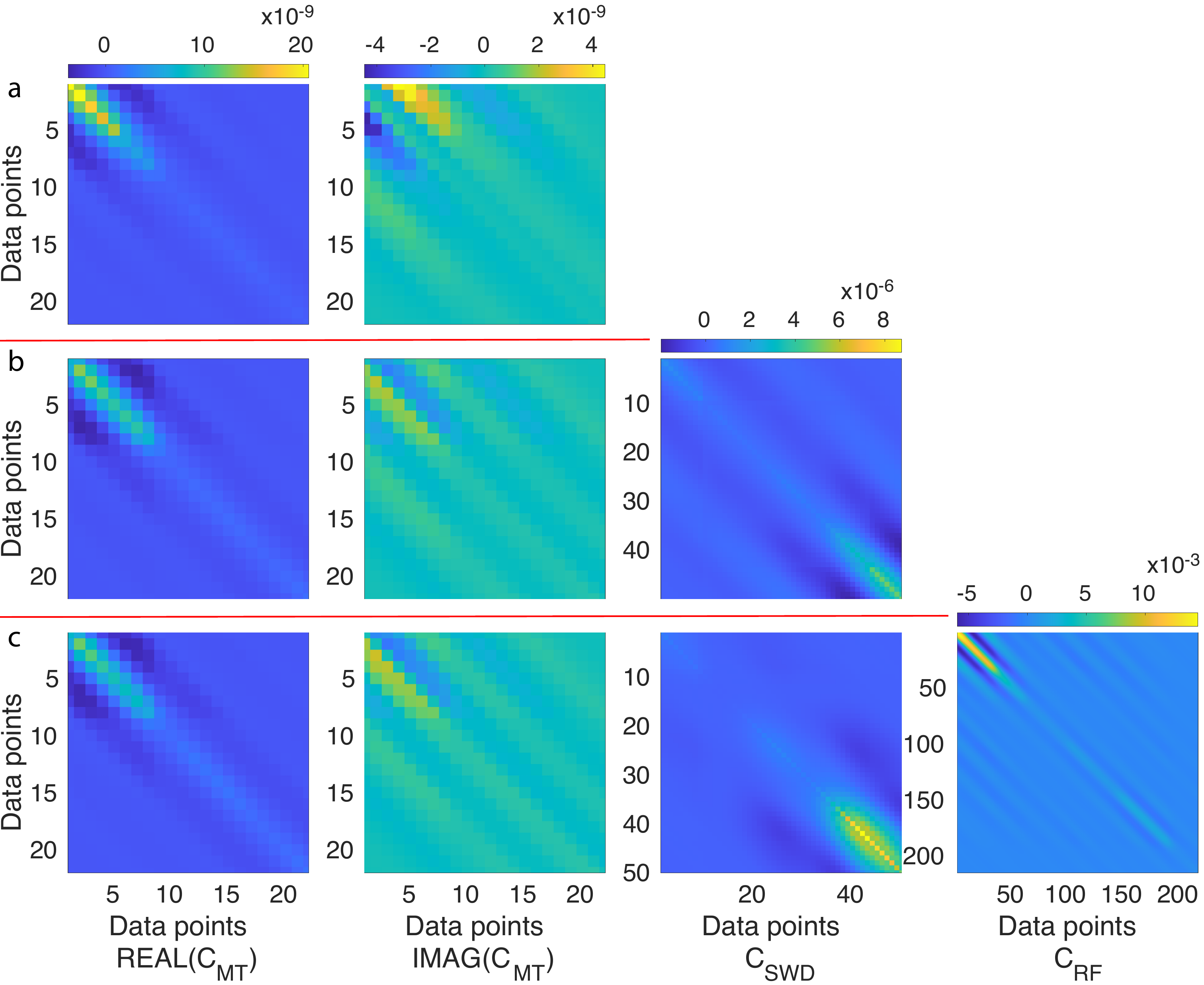}
    \caption{Empirically estimated data covariance matrices for (a) only MT, (b) MT-SWD, and (c) MT-SWD-RF inversions.}
    \label{fig-atha-cov}
\end{figure}

The estimated data covariance matrices for all cases are shown in Fig.~\ref{fig-atha-cov}. It is evident that all data types exhibit strongly correlated errors and clear non-stationary effects. Notably, the covariance estimate for RF data has the strongest correlations near the P-wave arrival and then at later times where multiples are typically found. This is because the RF shows a significant shift on the P wave that our model cannot fully explain. Furthermore, correlations at later times indicate that the predicted multiples are somewhat inconsistent with the observed multiples. 

\section{Discussion and Conclusions}
Our PMPI method has been successfully applied to both the simulated and field data. For simulated data, the various data types are intrinsically consistent. Hence, all data types are adequately fit, and uncertainties are reduced as additional data are introduced. This suggests that the MT, SWD, and RF data are complementary as long as there are no significant theory errors. However, as expected, complete consistency is not always the case for field data. 

Our field data includes three data types with minor inconsistencies. For example, the RF observation site in our study is separated from the MT and SWD site by $\sim$100 km which is a common limitation for 1D MPI~\citep[e.g.,][]{moorkamp2010joint}. Furthermore, the SWD data are obtained from a regional velocity model and not by direct processing of seismic waveforms. Such limitations can explain the increased uncertainty in the number of layers for all parameters in case 3 for the Athabasca data compared to the other cases (Fig.~\ref{fig-atha-k-sig}).

Figure~\ref{fig-atha-postpred}c shows that the inferred velocity profiles for case 3 do not explain the negative RF phase at $\sim$2 s and some phases between $\sim$6 to 11 s. The presence of negative crustal velocity contrasts in this area have been reported in other studies based on receiver functions~\citep{chen2015crustal} and reflection-seismic images~\citep{ross1997winagami,welford2006three}. However, the SWD data show no evidence of such structure and therefore discourage their appearance in our results. This is due to the fact that the regional velocity model from which the SWD data were derived does not contain the local features that may be present in the RF data(Fig. S15). Similarly, the MT data show no evidence of such structure, suggesting that if these structures exist, they do not exhibit changes in resistivities. The inferred values of the noise scaling parameter and the estimated covariance matrices (Figs~\ref{fig-atha-k-sig} and~\ref{fig-atha-cov}) show that the algorithm has assigned low error values ($\sim 0.05\%$) to the SWD data in cases 2 and 3. This is due to the lack of noise in the SWD data, a result of their derivation. This low noise level caused the SWD data to weigh significantly compared to the RF data in Case 3. This effect can also suppress local features that may be present in the RF data. Therefore, we interpret the results of these PMPIs as regional. The reductions in uncertainty observed in progressing from case 1 to case 2 and finally case 3, is robust since good data fits are achieved. Furthermore, our method provides flexibility to address minor inconsistencies by using advanced data covariance matrix estimation with hierarchical scaling. 

An important feature of our PMPI method is the variable complexity algorithm. This feature enables discontinuities for the various physical parameters that can be shared or distinct based on data information and was demonstrated with both simulated and field data. For the field data, the PMPI inferred two resistivity discontinuities (WCSB and the LAB) which is consistent with previous 2D and 3D MT inversions~\citep{wang2022three, boerner2000synthesis}. Although the estimated SWD noise is also orders of magnitude smaller than the MT noise, the algorithm did not force the resistivity structure to strictly follow the velocity structure (Fig.~\ref{fig-atha-marg}). Figure~\ref{fig-atha-postpred} shows that MT impedances are better fit for longer periods in all cases. This is most likely due to increased theory errors caused by local structure that is prominent at shorter periods/shallower penetration depths.

Our main result, PMPI case 3, infers four main regional features: the WCSB, a discontinuity in the crystalline basement, the Moho, and the LAB. We estimate the thickness of the WCSB in the range 0.88-1.4 km and the discontinuity in the crystalline basement to be located in the depth range 7.9-8.57 km. These values are consistent with other studies in the area~\citep{wang2022three,boerner2000synthesis,li2022earthquakes}. The Moho is estimated at 35.5-km depth with uncertainty between 32 km to 39 km. These values are in agreement with other studies, including studying RFs across Alberta~\citep[fig.~5]{chen2015crustal} and maps of the Moho depth across the North American Craton from six different seismic models~\citep[fig.~5]{gu2018precambrian}. Previous MT and seismic studies suggest average LAB depths of $\sim$200 km in the Athabasca area~\citep{wang2022three,bao2014plateau,chen2017finite}. Our PMPI method infers a LAB depth of 168.0 km with uncertainty between 155.0 and 196.0 km.

Most previous studies are based on linearized and regularized/smoothed inversions that make subjective choices. For example, to interpret the Moho in smooth models, a specific $V_s$ value is chosen ~\citep{chen2015crustal}. Similarly, when using MT the LAB depth may be identified based on a specific resistivity value ~\citep{wang2022three}, and with seismic interpretation the LAB location may be based on the center of the region with the strongest negative gradient in $V_s$~\citep{chen2007new}. Such linearized approaches can cause issues for robust uncertainty quantification. Our PMPI does not require subjective choices and discontinuities are resolved based on data information. In addition, subjective data weights are not required. Instead, the noise treatment provides weights based on data information. Finally, we note that MT-SWD PMPI provides most of the advantages in uncertainty reduction that MT-SWD-RF PMPI provides. Considering that RF data require long station records and are not available in all regions, MT-SWD PMPI can be an attractive, straightforward method for obtaining low uncertainty constraints on the lithosphere.

\section{Data availability}
Seismic waveforms for receiver functions can be downloaded from the Incorporated Research Institutions for Seismology (IRIS) website (https://ds.iris.edu/wilber3/). MT and SWD data are available upon request to the corresponding author and confirmation of coauthors.

\section{Acknowledgment}
All processing steps for the receiver functions prior to the deconvolution were done by using the Obspy package~\citep{beyreuther2010obspy} (https://github.com/obspy/obspy). The water level deconvolution and the forward model calculations for receiver functions were implemented by the Raysum package~\citep{frederiksen2000modelling} (https://home.cc.umanitoba.ca/~frederik/Software/). This research was funded by a Natural Sciences and Engineering Research Council of Canada Discovery Grant to Jan Dettmer.
%!!
%========================================================================================
%!!
%\bibliographystyle{gji}
%\bibliography{bibli}

\label{lastpage}
\end{document}